\documentclass[reprint,aps,prb,twocolumn,floats,showkeys,footinbib,superscriptaddress]{revtex4-1}
\usepackage[T1]{fontenc}
\usepackage{bm}
\usepackage{graphicx}
\usepackage{amssymb}
\usepackage{leftidx}
\usepackage{upgreek}
\usepackage{siunitx}
\usepackage{amsfonts}
\usepackage{amsmath} 
\usepackage[outdir=./]{epstopdf}
\usepackage{color}
\usepackage{float}
\usepackage{verbatim}
\usepackage{hyperref}
\usepackage{amssymb}
\usepackage{wasysym}
\usepackage{xspace}
\usepackage{ bbold }

\newcommand{\X}{X\xspace}

\newcommand{\Xp}{X$^+$\xspace}

\newcommand{\Mn}{Mn$^{2+}$}

\begin{document}

\title{\large \textbf{Impact of the Hole Gas on Optically Detected Magnetic Resonance in (Cd,Mn)Te Based Quantum Wells}} 

\def \FUW{Institute of Experimental Physics, Faculty of Physics, University
of Warsaw, ul. Pasteura 5, 02-093 Warsaw, Poland}

\def \MUN{Institute of Physics, University of Münster, 48149 Münster, Germany}

\author{A. \surname{{\L}opion}}\email[Corresponding author:]{a.lopion@uni-muenster.de}\affiliation{\FUW}\affiliation{\MUN}
\author{A. \surname{Bogucki}}\affiliation{\FUW}
\author{M. \surname{Raczyński}}\affiliation{\FUW}
\author{Z. \surname{Śnioch}}\affiliation{\FUW}
\author{K.\,E. \surname{Po{\l}czy\'{n}ska}}\affiliation{\FUW}
\author{W. \surname{Pacuski}}\affiliation{\FUW}
\author{T. \surname{Kazimierczuk}}\affiliation{\FUW}
\author{A. \surname{Golnik}}\affiliation{\FUW}
\author{P. \surname{Kossacki}}\affiliation{\FUW}

\begin{abstract}

 Optically detected magnetic resonance (ODMR) is a useful technique for studying interactions between local spins (magnetic ions) and carrier gas. We present the ODMR study of single {(Cd,Mn)Te/(Cd,Mg)Te} quantum wells (QWs) with the hole gas. We observe different characteristics of the ODMR signals obtained simultaneously using the optical signals of neutral and positively charged exciton. From that, we infer the existence of local fluctuations of carrier gas density resulting in separate populations of \Mn\ ions. At the same time, the shape of the ODMR signal contains information about the temperature of the magnetic ions involved in the absorption of the microwaves. Studying it in detail provides even more information about the interactions with charge carriers. In the QW, two separate ensembles of ions are thermalized differently in the presence of carriers.

\end{abstract} 
\keywords{ODMR, QW, spin-carrier interaction}

\date{\today}
\maketitle

\section{INTRODUCTION}

Effective coupling between magnetic ions and charge carriers is crucial for the realization of electrical control of magnetism, which is the foundation of the spintronics \cite{dietl2014dilute}. Cadmium manganese telluride is a perfect model system in which carrier density doping is independent of the concentration of incorporated magnetic ions \cite{gajIntroductionPhysicsDiluted2010}. In this work, we use the optically detected magnetic resonance (ODMR) technique as a very selective tool to study interactions between magnetic ions and charge carriers in the semimagnetic quantum wells (QWs) with hole gas. 

In the QW being the subject of the research presented here, it is possible to control the concentration of carriers by using the additional illumination \cite{kossackiNeutralPositivelyCharged1999, maslanaPtypeDopingII2003, kossackiOpticalStudiesCharged2003,kossackiPhotoluminescencePDopedQuantum2004,aku-lehMeasuringSpinPolarization2007}. Using the illumination above the energy of the bandgap of the barrier results in increase of the hole concentration. While the sample is illuminated only below the barrier bandgap energy -- the quantum well is more neutral. This effect can be observed in both photoluminescence and reflection measurements. This additional above-bandgap illumination is usually realized with the excitation laser for the photoluminescence study. During reflectivity measurements, apart from the incident light in the range corresponding to exciton structures, the sample should be illuminated additionally above the barrier energy gap -- to increase the carrier density in the QW. Moreover, the carrier concentration can be estimated based on the ratio of the amplitudes of the spectral lines corresponding to the neutral \X\, and the charged \Xp\, exciton present in the optical spectrum measured in a zero magnetic field. That allows one to simultaneously change and probe the carrier concentration in the nanostructure on an optical basis.

The ODMR technique is a selective tool to study various properties of the magnetic ions placed in the crystal lattice. Using ODMR studies, we can investigate the local strain or spin-lattice relaxation time \cite{baranov2008evidence,gurinODMREvidenceElectron2015, lopionMagneticIonRelaxation2022}. In this work, we show that the shape of the ODMR signal can also provide information about the carrier density interacting with the ions. 

\section{Sample and characterization}
\begin{figure}
    \centering
    \includegraphics{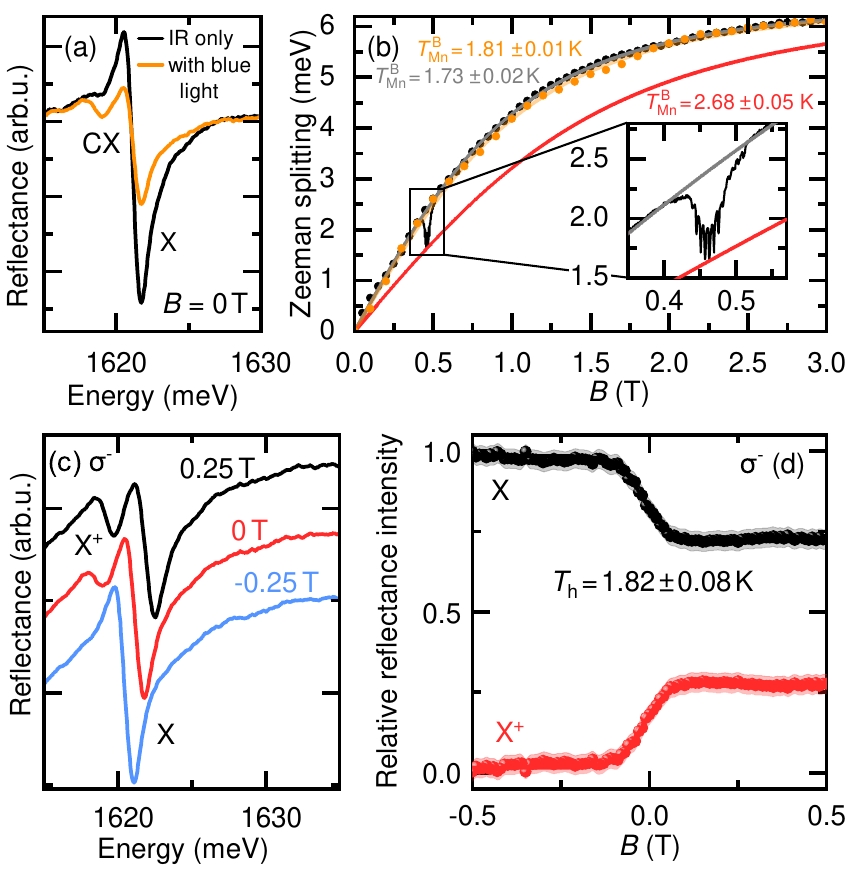}
    \caption{(a) Difference in the reflection spectrum in zero magnetic fields, in the absence and the presence of additional illumination with a blue LED. Two excitonic features are marked -- neutral exciton (\X) and positively charged exciton (\Xp). (b) Zeeman splitting of the neutral exciton line for the unilluminated situation (black points) with the fitted modified Brillouin function (gray curve) and the sample subjected to additional illumination (increased carrier density) - yellow points and yellow curve, respectively. The temperature of the ion system was a fitting parameter and was set to 1{.}73\,K in the non-illuminated situation and 1{.}81\,K in the illuminated situation. The black curve represents the ODMR signal obtained without the blue LED backlight for the frequency of microwave radiation $f$\,=\,12{.}9\,GHz, the red curve symbolizes the modified Brillouin function for the remaining parameters exactly as in the case of the gray curve but the temperature set to 2{.}68\,K. Inset: enlarged view of the region showing the detailed ODMR signal. (c) The dependence of the reflection spectrum for different magnetic fields with increased carrier density; (d) Relative intensity of the lines of the neutral \X\, and positively charged exciton \Xp depending on the magnetic field with fitted relationships -- according to the formula \eqref{eq:pol_hole}.}
    \label{pol_holes1}
\end{figure}

The studied structure was single 10\,nm (Cd, Mn)Te QW grown by molecular beam epitaxy (MBE) between symmetrical 50-nm (Cd,Mg)Te
barrier layers with Mg content of about 20\,\%. It was prepared on GaAs substrate with a thick buffer containing (3500\,nm) CdTe layer with an additional 2000\,nm (Cd,Mn)Te layer (Mn content the same as in the barriers). The Mg content in the barrier layer was determined by the reflectance measurements from the bandgap energy of the barrier \cite{waagGrowthMgTeCd1993}. The Mn$^{2+}$ content in the QW material close to 0.3\,\% was chosen to ensure sufficiently large Zeeman splitting while preserving narrow excitonic lines. The exact value was confirmed by fitting the Brillouin function \cite{gajMagnetoopticalStudyInterface1994} to the exciton Zeeman shift obtained from reflectance measurements in a magnetic field, see Fig.\ref{pol_holes1}(b).

Generally, (Cd,Mn)Te/(Cd,Mg)Te QWs are naturally p-type, even without intentional doping. In such a case, hole gas originates from background doping of the barrier and/or from the surface states, which can be especially influential for the QW placed near the surface \cite{maslanaPtypeDopingII2003}. Indeed, the presented samples contain hole gas in the QW. The sign of the carriers can be confirmed by using measurements of the photoluminescence signal in the magnetic field and observation of the spin singlet-triplet transition \cite{kossackiPhotoluminescencePDopedQuantum2004,lopionChargedExcitonDissociation2020}. The same group of samples was presented in Ref. \onlinecite{lopionChargedExcitonDissociation2020}, where their optical properties (along with the field of the spin singlet-triplet transition) were studied as a function of external illumination.

Here, we also used external illumination to increase carrier density in the samples. In this case, a blue LED (wavelength approx. 405\,nm) was used for providing such additional illumination. The LED was placed in front of the optical cryostat where the samples were located. With the presence of an additional illumination above the gap energy, apart from the line of the neutral exciton \X, we can observe an additional line in the reflection spectrum associated with the positively charged exciton \Xp, Fig. \ref{pol_holes1}(a). Based on the relative intensities of the lines in the zero magnetic field \X\, and \Xp, the density of holes is estimated to be below 1$ \cdot10 ^{10}$\,cm$^{-2}$ (without additional illumination) and 3\,$\cdot 10^{10}$\,cm$^{-2}$ (with additional illumination). In this paper, the two levels of carrier density are denoted as low and increased, accordingly.  

In the magnetic field, both exciton lines show a giant Zeeman shift. This shift is proportional to the magnetization of the \Mn system, as probed by band carriers (\cite{gajIntroductionPhysicsDiluted2010, gajRelationMagnetoopticalProperties1993}). It is well described by the modified Brillouin function, which exhibits a strong dependence on the temperature. Thus, one can consider how introducing carriers into the system affects the thermalization of magnetic ions excited by microwave radiation in the ODMR experiment. Figure \ref{pol_holes1}(b) shows the measurement of giant Zeeman splitting as a function of the magnetic field with low carrier density (without additional illumination). The modified Brillouin function was fitted to the splitting of exciton lines \X\, in a magnetic field. From that fit, the temperature of the system of polarizing ions is estimated at 1.73\,K. An analogous measurement was carried out for the increased carrier density -- the effective temperature of the ions increased slightly in this situation, up to 1.81\,K.

Figure \ref{pol_holes1}(c) shows reflection spectra measured with the increased carrier density for different magnetic fields. In one circular polarization, the \Xp line disappears. The positively charged exciton consists of two holes with opposite spins. The optical transition \Xp associated with the photocreation of a hole with a specific spin depends on the prior existence of a hole with an opposite spin \cite{huardBoundStatesOptical2000}. Hence, the oscillator strength of the charged exciton \Xp is proportional to the concentration of holes in the opposite spin subband. By tracking the changes in the relative intensity of the reflection lines \X\, and \Xp, it can be seen that in this case, the hole gas is completely polarized already in the field of about 0.1\,T, Fig. \ref{pol_holes1}(d). The following function, derived from Boltzmann distribution of holes between two spin levels, was fitted to the intensity changes in the field presented in the graph:

\begin{equation}
A_\pm (\Delta_{v,Z}) = \frac{A_\mathrm{max}}{1+\exp \left (\frac{\pm \Delta_{v,Z}}{k_\mathrm{B}  \mathrm{T}_\mathrm{h}} \right)},
\label{eq:pol_hole}
\end{equation}

where $A_\mathrm{max}$ is the maximum intensity of a given line, $\Delta_{v,Z}$ giant Zeeman splitting of the valence band, $T_\mathrm{h}$ is the carriers' temperature. This results in a temperature of the carriers slightly higher than the temperature of the ion system without microwave radiation. In the present paper, all the ODMR measurements were conducted in the magnetic fields above 0.1\,T, thus in the fully polarized hole gas regime. In the next parts, we show that the carrier gas stays in this regime even after the absorption of the MW radiation during the ODMR experiments.

\section{ODMR - setup and method of the measurements}
The transitions between Mn$^{2+}$ energy levels due to absorption of the applied microwave (MW) radiation lead to a decrease in magnetization. This can be evidenced by the decrease of the giant Zeeman shift. The effect can also be expressed in terms of increased effective spin temperature, which will be discussed later in this work. All the ODMR measurements presented in this work were conducted in the pulsed mode of microwave radiation and optical excitation to minimize non-resonant thermal drifts. Here, we define the ODMR signal as the difference in the energy position of the excitonic line in the reflectance spectrum for microwave and light pulses overlapping (ON) and shifted (OFF). The duration of the microwave pulse is about a few milliseconds. A representative ODMR signal as a function of magnetic field (ODMR spectrum) for fixed microwave frequency is present in Fig. \ref{pol_holes1}(b). Careful study of ODMR spectra can provide information about strain, what was discussed in detail by \textcite{boguckiAngleresolvedOpticallyDetected2022}. Moreover, the ODMR spectrum can provide information about the impact of interactions with carriers, which is the core of the work presented in this paper. 

The amplitude of the ODMR signal (the shift of the excitonic line) corresponds directly to the temperature of the magnetic ions, which creates net magnetization. After the sample is excited with pulsed microwave radiation of frequency of 12{.}9\,GHz, a decrease in Zeeman splitting for a resonant magnetic field of 0{.}46\,T can be observed, see Fig.\ref{pol_holes1}(b). This reduction in Zeeman splitting corresponds to an increase in the temperature of the ion system to 2{.}7\,K as shown by the red curve in the figure.

An important detail is that the measurement conditions are set to avoid non-resonant signals. It can be evidenced by the fact that outside the resonance pattern, the position of the excitonic line with and without the microwaves is the same -- see inset of Fig. \ref{pol_holes1}(b). During the experiment, the overall temperature increase results from the resonant and non-resonant heating of the magnetic ions system. The measurements were conducted in the pulsed mode, in the timescale where we can assume that the non-resonant heating is constant. In Fig. \ref{power_dep}(a), the ODMR spectra obtained with increasing microwave excitation power are presented. The signals reveal two distinct components: the resonant ODMR spectrum, which appears within a specific range of magnetic fields where resonance conditions are met, and a non-resonant signal that persists across the entire range of magnetic fields. This non-resonant offset is attributed to the non-resonant heating of the magnetic ion system via the crystal lattice or carrier gas. In Section \ref{ESTI}, we show that adding carriers to the sample (e.g., by additional illumination) impacts non-resonant heating in the presence of microwave radiation. Thus we can assume that most of the non-resonant heating is connected to the absorption of the MW in the carrier gas.  

The excitation microwave power input can be directly connected to the magnetic ions' temperature derived from the Zeeman splitting, as shown in Fig. \ref{power_dep}(b). The input power was set to -10\,dBm for further microwave radiation measurements to minimize the non-resonant heating. For such an excitation power, the ODMR signal is relatively good, and the temperature derived from the Zeeman splitting (1.81$\pm$0.01\,K) is only slightly elevated in comparison to the temperature without microwaves (1.73$\pm$0.01\,K), Fig. \ref{pol_holes1}(b).

\section{Shape and position of the ODMR signal}
The ODMR spectrum has a specific shape governed directly by the structure of Mn$^{2+}$ energy levels. They can be described by the spin Hamiltonian\cite{qazzazElectronParamagneticResonance1995, boguckiAngleresolvedOpticallyDetected2022,abragamPrinciplesNuclearMagnetism1961}:

\begin{figure}
\centering
\includegraphics{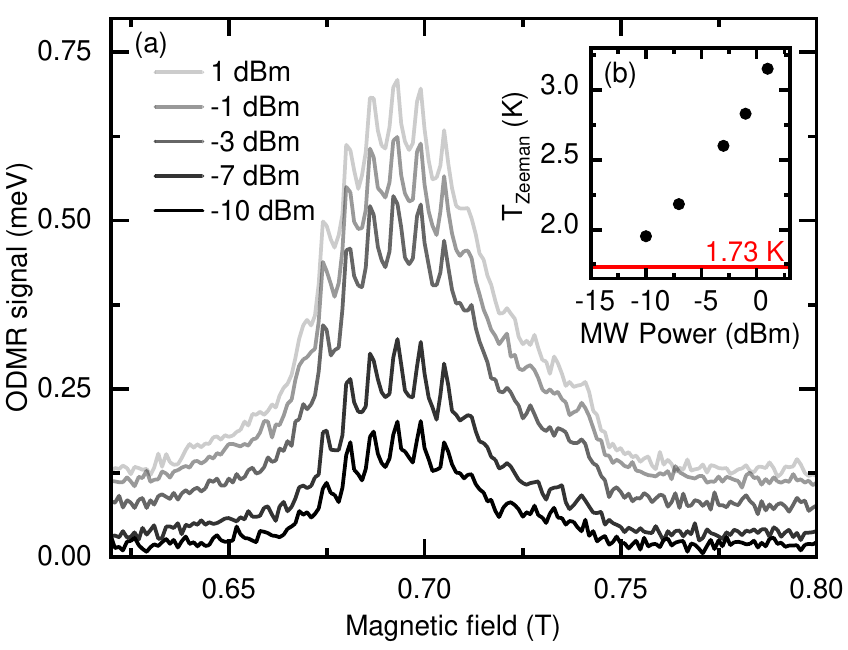}
\caption{(a) ODMR signal power dependence, both amplitude of the signal and the offset are increasing with the applied microwave power (b) Temperature derived from Zeeman splitting change vs. microwave excitation power dependence. Full Giant Zeeman splitting for this sample, bath temperature, and the magnetic field are about 3\,meV. That corresponds to the temperature marked with a red line -- derived from Zeeman splitting measured without microwave radiation. The microwave excitation power in the ODMR measurements was set to -10\ dBm to give a significant ODMR signal and avoid non-resonant signals, which are represented by an increase in the offset.}
\label{power_dep}
\end{figure}

\begin{figure}
\centering
\includegraphics{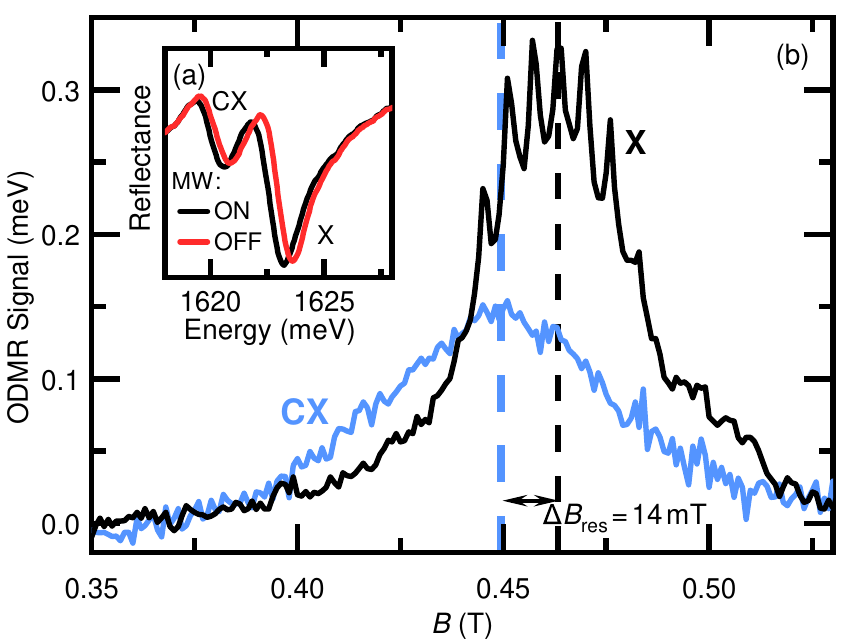}
\caption{(a) Reflection spectrum of the sample measured at temperature $\sim$1.6\,K in magnetic field $B=0.463$\,T with microwave radiation resonant for this field (12{.}9\,GHz) at in the "ON" and "OFF" states, the change of position is visible on both visible exciton complexes; (b) The ODMR signal as the difference in the positions of the energy lines observed in the reflection spectrum in the presence and absence of microwave radiation at the frequency $f$\,=\,12{.}9\,GHz depending on the value of the applied external magnetic field; the signal observed on the line of the charged exciton \Xp is shifted ($\Delta B_\mathrm{res}$\,=\,14\,mT) towards lower magnetic fields compared to the signal obtained on the neutral exciton \X. }
\label{Knight_UW1467}
\end{figure}

\begin{multline}
H_\mathrm{Mn^{2+}} = g_\mathrm{Mn} \mu_\mathrm{B} \boldsymbol{\mathrm{B}} \cdot \widehat{\boldsymbol{\mathrm{S}}} +  A \widehat{\boldsymbol{\mathrm{I}}}  \cdot \widehat{\boldsymbol{\mathrm{S}}} + \\ +  \frac{a}{6} \left ( \widehat{S}_x^4+\widehat{S}_y^4+\widehat{S}_z^4- \frac{S(S+1)(3S^2+3S-1)}{5}\right) + \\ +  D \left( \widehat{S}_z^2 - \frac{S(S+1)}{3} \right),
\end{multline}

where $\widehat{\boldsymbol{\mathrm{S}}}$ and $\widehat{\boldsymbol{\mathrm{I}}}$ are shorthand for $\widehat{\boldsymbol{\mathrm{S}}}\otimes\mathbb{1}$,$\mathbb{1} \otimes \widehat{\boldsymbol{\mathrm{I}}}$
respectively. The first term of the Hamiltonian describes the Zeeman splitting of the energy levels in a magnetic field due to the electron d-shell spin ($S$). $g_\text{Mn}$ is the g-factor of the manganese ion. The second term describes hyperfine splitting, where $\widehat{\boldsymbol{\mathrm{I}}}$ denotes the manganese ion nuclear spin operators. \Mn\, ion
has electronic spin S\,=\,5/2 and nuclear spin I\,=\,5/2.  The third term for the level-splitting part of the Hamiltonian is due to the placement of the ion in the crystal field with the symmetry T$_\mathrm{d}$.

The last term refers to the zero-field splitting related to lowering the symmetry of the crystal and corresponds to the uniaxial stress associated with deformation in the $z$ direction. The strain-induced axial-symmetry parameter $D$ is a component characteristic for a given sample, resulting from the parameters of growth, composition, and thickness of individual layers. $D$ can be determined from the mismatch of the material lattice constants of the quantum well $a_\mathrm{QW}$ and the buffer $a_\mathrm{buf}$ by determining the strain $\varepsilon = \frac{a_\mathrm{QW}-a_\mathrm{buf}}{a_\mathrm{QW}}$. Finally, the $D$ parameter \cite{qazzazElectronParamagneticResonance1995} can be determined: $D=-\frac{3}{2} G_{11} \left (1+ \frac{2C_{12}}{C_{11}} \right) \varepsilon$, where $G_{11}$\,=\,$72{.}2$\,neV  and $\frac{2C_{12}}{C_{11}} =0.7$\cite{boguckiAngleresolvedOpticallyDetected2022,greenoughElasticConstantsThermal1973}.

The electronic and nuclear spin coupling for S and I equal to 5/2 results in 36 energetic states of Mn$^{2+}$ ion. In a zero magnetic field, the splitting of those states corresponds mainly to the strain ($D$ parameter). If the magnetic field is non-zero but small (B<0.05T), the evolution of the states is complicated, and the eigenstates of the Hamiltonian are still not eigenstates of the $S_z$ operator. The Zeeman term dominates Hamiltonian for higher magnetic fields, and the energy levels begin to have a well-defined spin projection to the magnetic field axis ($z$ direction). Considering the selection rules for magnetic dipole transitions ($\Delta S_z = \pm 1$, $\Delta I_z = 0$), the ODMR spectrum contains 6 sets with 5 transitions in each.     

The shape of the ODMR spectrum depends on the separation and the population of energy levels participating in microwave absorption. The intensity of a single transition corresponds to the occupation of the states between which the transition occurs—thus, it is strictly connected to the temperature of the magnetic ions. Qualitatively, the features observed at the higher magnetic field correspond to the transitions between lower energy levels than features observed at the lower magnetic field.
 
In the case without additional illumination, the carrier density is low, and the neutral exciton transition dominates the reflectance spectrum. When additional illumination is set on the sample, the carrier density is increased. That is evidenced in the reflectance spectrum, which shows another excitonic transition in the energy below \X: charged exciton (\Xp). Both excitonic lines observed in the reflectance are sensitive to microwave radiation, Fig. \ref{Knight_UW1467}(a). We observe that both \X\ and \Xp lines move towards lower energies when the light and microwave pulses coincide in time ("ON" state) relative to the position of the lines when the light and microwave pulses are significantly temporarily separated ("OFF" state).

Here, we focus on the observed differences in the ODMR signal obtained as a change in the position of neutral and charged exciton optical features. Generally, the neutral and charged excitonic complexes have different natures and respond differently to experimental variables. However, in the ODMR measurements presented in this work, we can treat both excitonic features as probes for the system of magnetic ions, which is the main subsystem absorbing microwave radiation. With this approach, there should be no difference if the magnetic ions are examined using neutral, charged, or any other excitonic line, as the probed system (magnetic ions) is the same. However, this is not the case in the shown studies. We can observe differences in the ODMR signal obtained as the position change of the neutral excitonic feature and the charged one. Figure \ref{Knight_UW1467}(b) shows the ODMR signal measured as a shift of both \X and \Xp lines. In the ODMR signal measured as a change of the position of the neutral exciton line, many narrow features can be seen. Those are related to the transitions between different states of the manganese ion split by the magnetic field, hyperfine interaction, and strain. At the same time, we notice that such a structure is not visible in the signal monitored on the charged exciton line, and the shape of the signal seems smoother. Moreover, the maximum of the ODMR signal for \Xp line is shifted relative to the maximum of the signal measured on the neutral exciton towards lower magnetic fields. These differences suggest that the two lines probe different populations of \Mn\, ions, which exhibit different paramagnetic resonance broadenings and shifts. 

In the next parts, we show that the resonance shift observed on the charged state line results from interactions between magnetic ions and charge carriers. The $\Delta B_\mathrm{res}$ is connected to the density of carriers. Here, and in a similar situation in the following parts of the paper, the center of the resonance for the excitonic state is calculated as the center of mass of the ODMR spectrum (ODMR signal vs. magnetic field). That approach gives us precise information about the position of the resonance when it is more complicated in shape. However, it is important for the calculation to subtract the offset of the ODMR spectrum correctly. A similar result can be obtained by fitting a peak function (e.g., Gaussian) to the data.  

It should be noted here that both signals are acquired simultaneously, as they are extracted from the same optical spectrum. The ODMR signal obtained by observing the \X line appears unaffected by the interaction with the charge carriers, while that obtained by observing the \Xp line is apparently modified. This result suggests that the absorption associated with the neutral exciton occurs in different, spatially separated regions than the absorption associated with the charged exciton. At the same time, the spot of light used for reflectance measurements covers and probes both types of areas of different local carrier densities. The sample is relatively homogenous regarding the optical response in the scale above the laser spot size ($\sim$100 $\mu m$). That suggests the discussed inhomogeneities of the carrier density distribution are relatively small, below the spatial resolution of this optical measurement.

\section{Estimation of effective temperature of magnetic ions \mbox{\Mn\, and carriers}}
\label{ESTI}

\begin{figure}
\centering
\includegraphics{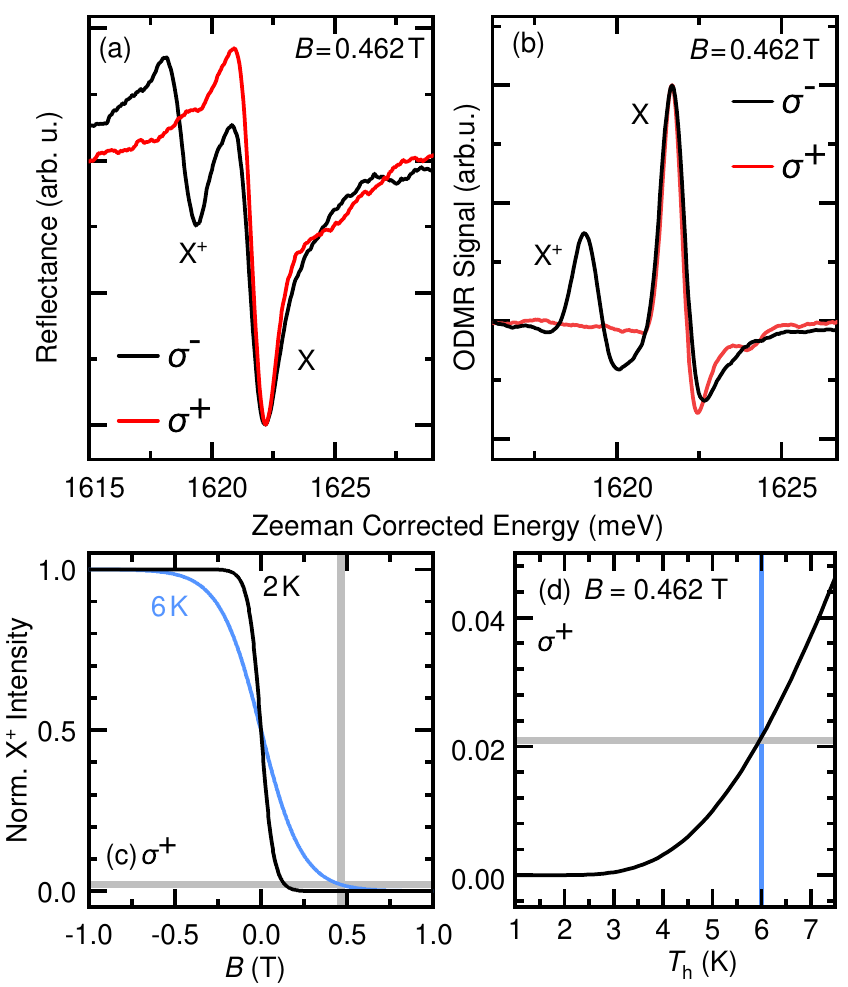}
\caption{(a) Low-temperature (helium bath temperature $T$\,=\,1{.}6\,K) reflection spectrum measured in the presence of additional backlight and microwave radiation (a microwave pulse overlapped with a light pulse) with frequency $f$ \,=\,12{.}9\,GHz in a resonant magnetic field $B$\,=\,0{.}462\,T measured in two circular polarizations; (b) Differential ODMR signal for measurement (a), (a,b) were corrected for giant Zeeman shift; (c) Curves representing the normalized signal of a charged exciton \Xp in a magnetic field, according to the formula \eqref{eq:pol_hole} -- carrier polarizability, for two different carrier gas temperatures 2\,K and 6\,K; (d) Dependence of the exciton signal \Xp on the temperature of the hole gas, according to the formula \eqref{eq:pol_hole}}
\label{eng_holes}
\end{figure}

We exploited the possibility of controlling carrier gas density with additional illumination to elucidate the effect of carrier concentration on the ODMR signal obtained on both excitonic features. The question can be asked whether applying microwave radiation to the sample is free of the undesired effects, such as the increase in the effective temperature of the carrier gas and the resulting depolarization of the carriers. Figure \ref{eng_holes}(a) shows the spectra of reflectivity in the presence of microwaves ($f$\,=\,12{.}9\,GHz) in a resonant magnetic field for two different circular polarizations, corrected for the giant Zeeman shift, so as to emphasize the difference in the intensity of the exciton structures. In the $\sigma^-$ polarization, there are two excitonic features -- while in the opposite polarization, we do not observe the lines of the charged exciton \Xp. To make it more pronounced, we can use the difference of the optical spectra measured in ON and OFF states -- such a difference we call differential signal. In a situation without microwaves and in the magnetic fields where the carrier gas is polarized (B>0.1\,T),  the charged exciton feature is present only in one circular polarization of light. When we apply microwave radiation, the polarization of the hole gas may change, resulting in a non-zero differential signal. However, the lack of reflection feature of the charged exciton \Xp in the $\sigma^+$ polarization proves that even in the presence of microwave radiation, the carrier gas is not heated sufficiently to depolarize its spin in the $B$\,=\,0{.}462\,T field. This gives us an upper estimate for the temperature of the carrier gas $T_\mathrm{h}$. Assuming that the uncertainty of the reflection measurement determines the minimum observable intensity \Xp (gray line in Figure \ref{eng_holes}(c), we can use the formula \eqref{eq:pol_hole} to find that the temperature of holes under the influence of microwave radiation $T_\mathrm{h} < 6$\,K, Fig. \ref{eng_holes}(d). Whether we are dealing with a gas that is fully or only partially polarized also depends on the value of the magnetic field. However, in most of the measurements presented in this paper, the smallest magnetic fields of the ODMR measurement had a value of about 0.4\,-\,0.5\,T. Hence, it seems reasonable to assume that in all cases, we are dealing with a polarized hole gas, even when the carrier gas temperature is increased due to the subjecting of the sample to microwave radiation.

\begin{figure}
    \centering
    \includegraphics{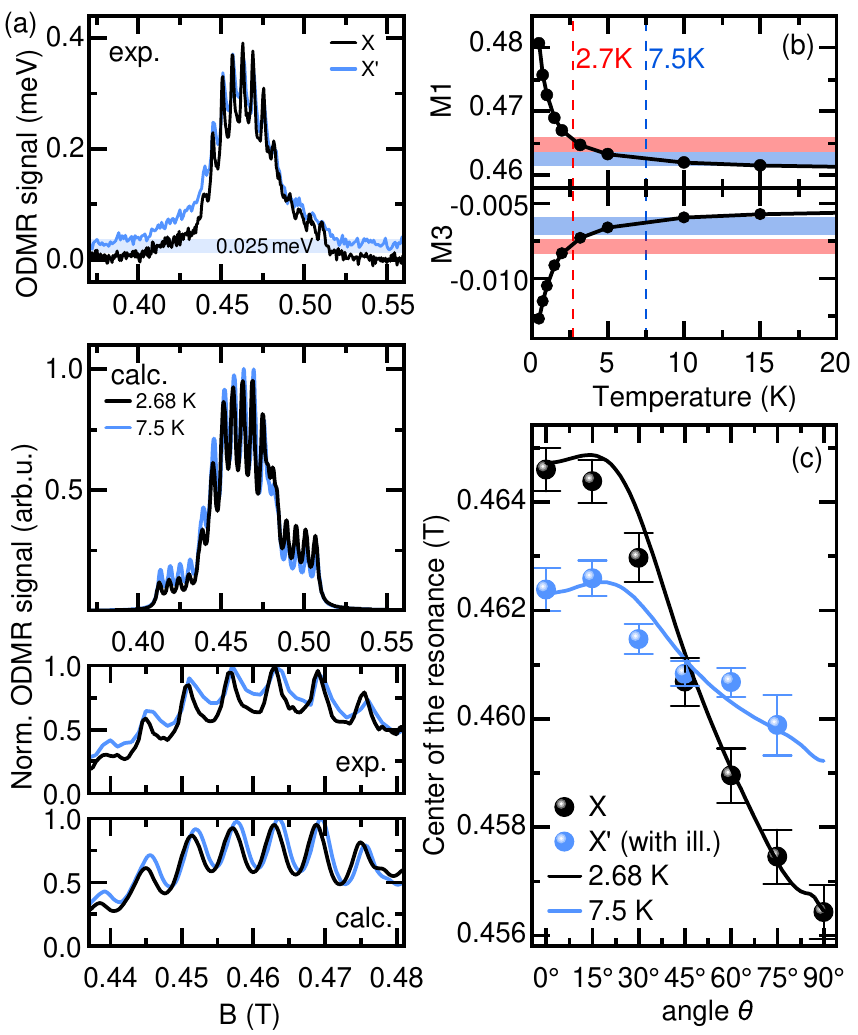}
    \caption{(a) ODMR signal as a shift of neutral exciton, with MW frequency $f$\,=\,12{.}9\,GHz measured without (\X) and with increased carrier density (\X'), helium bath temperature 1{.}6\,K; numerical calculation of ODMR spectra for $D$\,=\,780\,neV, $C$\,=0 and Mn ion temperature $T$\,=\,2.68\,K i $T$\,=\,7.5\,K. In the figure, the offset corresponding to non-resonant heating is marked. To calculate the central moments, it is crucial to subtract this part of the spectrum. (b) black connected points - change of first (M1) and third (M3) central moment in the function of Mn ion temperature calculated from numerical calculated ODMR spectra, thick blue and red lines represent M1 and M3 calculated for data measured for low (blue) and increased (red) carrier density and corresponding temperatures of Mn ions, marked with the dashed lines. (c) Center of the resonance for neutral exciton measured without (black spheres) and with increased carrier density (blue spheres) for different angles of the external magnetic field, colored curves mark the numerically calculated evolution corresponding to different temperatures of Mn ions.}
    \label{heating_deg}
\end{figure}

As discussed in the Introduction, the most straightforward way to determine the temperature of the Mn subsystem is fitting the Brillouin function to the observed Zeeman shift. However, we could also employ an alternative approach based on the results of the ODMR measurements. This method relies on the observation that the shape of the ODMR spectrum obtained by observing the neutral exciton with lower and higher carrier density is slightly different. Still, the position of the sharp transition lines is the same.  As such they can be therefore used to extract the temperature of the system. Quantitatively, the temperature value was extracted from numerical calculations of the whole ODMR spectrum. Figure \ref{heating_deg}(a) shows the comparison between measured and calculated ODMR spectra for the magnetic field perpendicular to the plane of the QW and two different illumination conditions, which corresponds to different temperatures -- it can be seen that they quite well reflect the differences observed in the spectra measured with low and increased carrier gas concentration. We also observe the emergence of an offset (highlighted in the blue region) corresponding to non-resonant heating when the sample is exposed to MW radiation in the charged regime. This offset is not present in the non-charged regime (black curve). This indicates that the non-resonant heating is primarily driven by the absorption of the microwaves by the carrier gas.

As mentioned before, a strict indicator of the temperature value is the shape of the ODMR spectrum. As the spectrum can be complex, we propose using the measure given by the data central moments. The first central moment (M1) provides information about the center of mass of the spectrum. The third one (M3) is connected to the asymmetry of the spectrum (with respect to M1). Both change with the temperature as the occupation of the manganese levels, and consequently, the transitions change the intensities, Fig. \ref{heating_deg}(b). The exact values of M1 and M3 change with the properties of the sample (strain, composition). However, the quantitative behavior with temperature is the same. For each sample and the given data set, it is much easier to calculate the moments and estimate the temperature of the magnetic ions on their basis.

\section{Dependence on magnetic field angle}

To further test our interpretation, we also performed measurements with varying angles of the external magnetic field. The shape of the ODMR spectrum obtained on the neutral exciton line was examined depending on the angle of the external magnetic field $\theta$ with respect to the growth axis. The zero angle corresponds to the measurement in the Faraday configuration (magnetic field perpendicular to the QW plane). The results obtained for the situation with low (\X) and increased carrier density (\X') are presented in the figure \ref{heating_deg}(c). Due to the increased carrier density, a significant change in the thermal distribution of manganese ion states leads to changes in the shape of the ODMR signal and an effective shift of their center. The dependence of the resonance field for two different temperatures, 2.68\,K and 7.5\,K, determined based on numerically calculated ODMR spectra, was plotted on the graph \ref{heating_deg}(b) obtaining a very good agreement with the experimental points. Such changes in the resonance field depending on the $\theta$ angle confirm that the temperatures of the ion system in the low and increased carrier density conditions are significantly different. Measurement of the ODMR spectrum and its center as a function of a magnetic field angle confirms the accuracy of the used model and the high temperature of the magnetic ions under MW radiation and with increased carrier density.


\begin{figure}
    \centering
    \includegraphics{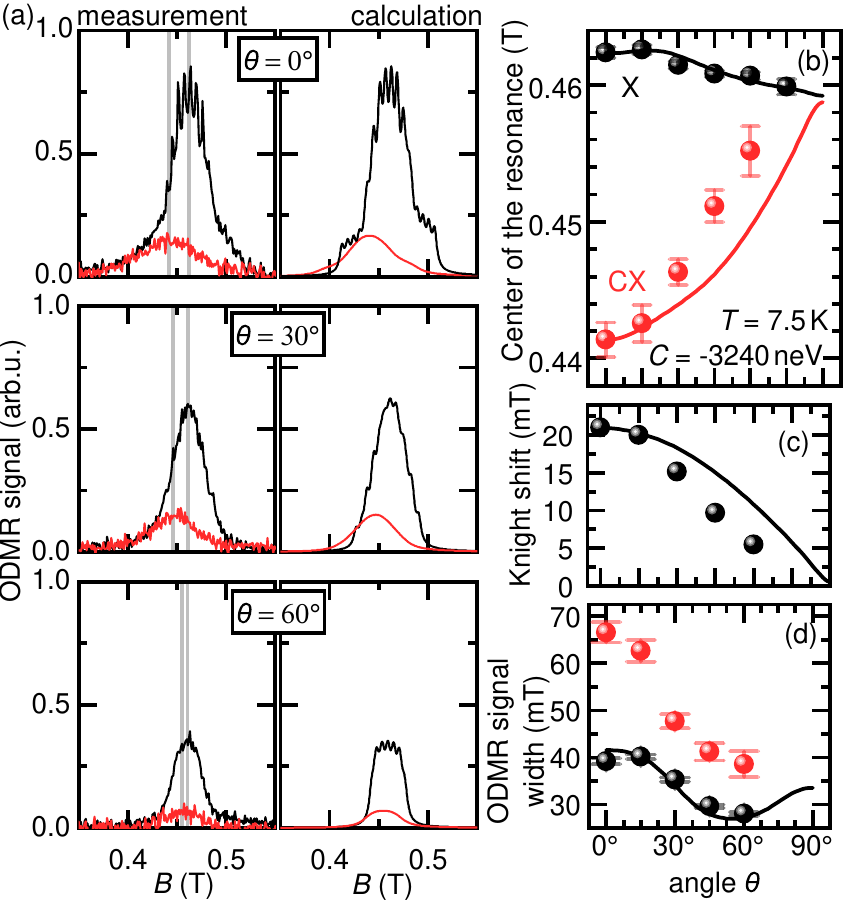}
    \caption{(a) Dependence of the ODMR signal for microwave frequency 12.9\,GHz observed by examining the position of the neutral (\X) and charged (\Xp) exciton lines on the magnetic field for different field angles relative to the samples plane -- and simulated spectra ODMR for parameters: $T$\,=\,7{.}5\,K, $C$\,=\,0 (\X) and $C$\,=\,-3240\,neV (\Xp) , $ D$\,=\,780\,neV. The constant $C$ is the carrier-ion interaction constant. (b) Positions of the ODMR signal maxima for the charged and neutral exciton depending on the angle of the magnetic field relative to the sample plane (0$^\circ$ means Faraday configuration, 90$^\circ$ -- Voigt); (c) The distance of the resonance magnetic fields measured by observing the lines \X\, and \Xp\, (Knight shift) depending on the angle of the magnetic field; (d) ODMR signal width measured by observing \X\, and \Xp, black curve represents simulated values.}
    \label{degree_045}
\end{figure}

Here, we show more differences in the ODMR signals obtained as a change of the excitonic line of neutral and charged exciton. Figure \ref{degree_045}(a) shows the ODMR signals depending on the magnetic field and for different angles of the field relative to the plane of the sample. We can observe that the distance between the ODMR signals for the two lines is different depending on the magnetic field angle. Both the ODMR signal obtained by observing the \X\ and \Xp\ lines change the location of the center of gravity with the change of the direction of the magnetic field, Fig. \ref{degree_045}(b), and their mutual distance is decreasing. 

This was another observation that suggests a difference in the properties of magnetic ions probed with the neutral excitonic feature to those probed with the charged excitonic line. In the next part, we show that the characteristics of the ODMR signal obtained with the charged exciton line can be connected directly to the interactions between probed magnetic ions and charge carriers. 

For the interactions with the carriers (hole gas), a simple Heisenberg model was used, adding to the Hamiltonian mean-field exchange term:
\begin{equation}
H_{p-d} = - \beta n |\varphi(z)|^2 \langle \hat{s} \rangle \cdot \hat{\mathrm{S}}_{Mn} = C \langle \hat{s} \rangle \cdot \hat{\mathrm{S}}_{Mn}
\label{eq:c}
\end{equation}

where $\beta$ is exchange integral between hole and Mn ion, $n$ -- carrier density, $|\varphi(z)|^2$ - distribution of carriers along growth axis $z$.

In this model, we assume that localized magnetic ions in the QW interact with fully polarized hole gas. Two facts justify this assumption. First, from the hole polarizability measurements, we observe that even if the carrier gas temperature is elevated after the absorption of the MW, the carrier gas stays in the fully polarized regime. Additionally, we do not detect any ODMR signal shift towards higher magnetic fields. Such a signal could be observed in the case of interaction with a hole spin anti-parallel to the ground spin state. In the fully polarized hole gas regime, the holes of the different (e.g., antiparallel) spin could be linked to strong localization and coupling with manganese ions. However, we have not observed such a signal in the experiment. That also confirms the high quality of the sample and low density of the localizing sites for carriers.   



Therefore, we use mean-field approximation, and the carrier's spin -- $\langle\hat{s} \rangle$  denotes the average over the whole gas of holes. The shift towards lower magnetic fields is connected only to the effective magnetic field derived from polarized holes. 

In the case of a magnetic field perpendicular to the plane of the well ($\theta = 0^\circ$), the shift of the resonances measured between both exciton complexes is about 23\,mT -- which is reproduced for the constant $C$\,=\,-3240\, neV. This value, according to the equation \eqref{eq:c}, corresponds to a hole gas concentration of 4.05\,$\cdot$10$^{10}$\,cm$^{-2}$. 

It was already shown for similar structures, in \onlinecite{kossackiNeutralPositivelyCharged1999,kossackiPhotoluminescencePDopedQuantum2004} that the mutual intensity of the features corresponding to the neutral and charged exciton gives information about the carrier density. On that basis, we estimated the carrier concentration to be about 3\,$\cdot 10^{10}$\,cm$^{-2}$ (with additional illumination). The value of carrier density obtained from the Knight shift is slightly higher than estimated from reflection spectra. Assuming that the carrier density obtained from the reflectance spectrum is a mean over the laser spot area and that the Knight shift is a local result for areas with the excess of carriers, we can estimate the percentage of those areas -- about 75\%.

The change of the center of gravity for ODMR spectra obtained as a change of neutral and charged exciton is shown in Fig \ref{degree_045}(c) along with the curve obtained as the difference between the modeled behavior of each ODMR spectrum shown in Fig \ref{degree_045}(b) with red and black curves. In Figure \ref{degree_045}(d), the changes in the width of the ODMR signal are also shown. Both of those characteristics -- the center of gravity and the width are derived from rather complex ODMR spectra. However, as was discussed in detail in the previous part for the neutral regime -- the central moments of spectra give a fingerprint of the samples' condition, i.e. the temperature of magnetic ions. Here again, the first moment M1 (center of gravity) and the second M2 (width) show the agreement between the experimentally obtained data and the modeled spectrum behavior in the magnetic field.   


\section{Discussion}

Even for such a simple toy model, the agreement with the experimental data is satisfactory. One of the possible reasons for the remaining deviations is the strong assumption of the equal temperature of the magnetic ions probed by \X\ and \Xp, or the parameters remain constant with changing magnetic field angle. The exchange integral in this system is isotropic. However, the $C$ parameter contains the local density of the carrier wavefunction, which, in contrast, can depend on the $\theta$ angle. We consider a uniform distribution of charges in the plane of the well. At the same time, the literature data \cite{golnikMicrophotoluminescenceStudyLocal2004,MaslanaMicrophotoluminescenceStudyPtype2006, lopionChargedExcitonDissociation2020} and the fact of simultaneous observation of different ODMR signals on the \X\, and \Xp\, lines show that this is not true. Most likely, in the plane of the well, there are areas with an excess of carriers, from which the optical signal comes from the recombination of charged excitons, and areas with a lower carrier density. In this case, the size of the areas is submicron, and in the measurement of photoluminescence or reflection in the macro scale, it is impossible to separate them.

The observation about the spatial separation of \X and \Xp goes along with the long-running discussion on that topic. Significant works, such as \onlinecite{moody2014coherent,brunhes1999oscillator}, have confirmed the correlations between charged and neutral excitons, which is serious evidence for the lack of (or significantly small) separation between the two. However, the level of the separation or its absence can be connected to the (in-) homogeneity of the Coulomb potential in the QW layer. This Coulomb disorder may be a result of the inhomogeneity of the surface of the sample for the systems with low separation between QW and the surface \cite{lopionChargedExcitonDissociation2020,MaslanaMicrophotoluminescenceStudyPtype2006} or inhomogeneity of the QW layer itself. In our case we believe it is a first case scenario -- as we observe the possibility of change of the Coulomb disorder with changing the additional illumination and screening the surface potential. 

In the next part, we discuss the temperatures of the subsystems that contribute to the spin relaxation processes observed with the ODMR signal. 

When the density of hole gas is low, the temperature of the magnetic ions derived from the giant Zeeman curve is similar to the temperature that can be derived from the ODMR measurement -- both from the intensity of the single MW transitions and the center of mass of the ODMR spectrum vs. magnetic field angle. 

For increased carrier density, the situation becomes more complex. In such a case, the two transitions visible in the reflectance spectrum -- neutral (\X) and positively charged exciton (\Xp) give different characteristics of the ODMR signal. In this case, we assume that the optical signal from charged and neutral exciton comes from spatially separated (in plane of QW) regions differing in the carrier density. In the regions with the higher density of carriers, where the optical signal of charged exciton comes from, the ions experience an effective exchange field of polarized hole gas, inducing ODMR signals shifting towards a lower external magnetic field. However, even the ions probing the "neutral" regions experience interactions with the increased density of carriers. In this case, the change in the ODMR spectrum is fine but significant. It becomes even more apparent when we observe the behavior of the center of mass of the ODMR spectrum with a magnetic field angle. The temperature of the magnetic ions estimated based on the ODMR measurements seems to be higher than the temperature derived from giant Zeeman splitting in the same conditions.  This means that these two measurements (ODMR and Zeeman splitting) examine different groups of ions. Part of the ions absorbing radiation microwave has a greater contribution to the ODMR spectrum and, simultaneously, a higher temperature than other ions, contributing mainly to building magnetization in the sample. In general, all ions in the QW should absorb microwave radiation the same way. However, there may be differences in rates of energy and spin transfer. A less effectively relaxing population would be favored in creating the ODMR signal. Due to increased interactions between them, the others would be less active in resonance studies, as they relax much faster and their lines are broadened.

\begin{table*}
\begin{center}
\begin{tabular}{l | c | c| c |c  } \hline
 \hline
& & & &\\
 & Brillouin & Brillouin & ODMR & hole polarizability ($T_\mathrm{h})$\\[1ex]
& ($T_\mathrm{Mn}^\mathrm{B}$) & ODMR ($\Delta$$T_\mathrm{Mn}^\mathrm{B}$) & shape & (without and with MW)\\[1ex]
\hline \hline
& & & &\\
Low hole density  & 1.73\,$\pm$\,0.02\,K  & 2.68\,$\pm$\,0.05\,K & 2.68\,K & -  \\[1ex]
\hline \hline
& & & &\\
Increased hole density\, \, \,  & 1.81\,$\pm$\,0.01\,K & $\sim$\,2.7\,K & 7.5\,K & 1.82/<6\,K \\[1ex]
\end{tabular}
\caption{Table summarizing the temperatures of different subsystems in the QW measured or estimated with different methods. The temperature of the helium bath in all cases was about 1{.}6\,K, and in the case of pulsed microwave excitation, it slightly increased by about 5-10\%. In order to be able to compare possibly similar measurement conditions in the table the microwave excitation means excitation with $f$\,=\,12{.}9\,GHz, $P_\mathrm{ MW}$\,=\,$-10$\,dBm. As the power emitted by the antenna and absorbed by the sample may differ for different frequencies, one should be careful to compare carrier or ion temperatures in such cases.}
\label{tab:temp_1}
\end{center}
\end{table*}

Previous works have shown that different subsystems present in the sample --  manganese ions, charge carriers, and crystal lattice -- can vary in effective temperature \cite{scherbakovAccelerationSpinlatticeRelaxation2001,konigEnergyTransferPhotocarriers2000, scherbakov1999heating}. Here, we deal with an even more complex situation in which manganese ions do not form a homogenous system. A summary of the temperature results obtained in this paper is provided in the table \ref{tab:temp_1}. 

If the system is not exposed to microwave radiation, the series of temperatures is as follows: the lowest temperature is the temperature of the helium bath $\sim$1.6\,K, then the temperature of the ions $T^B_\mathrm{Mn}$ obtained from Brillouin function fitted to magnetization in the magnetic field (without microwaves), the temperature of the holes $T_\mathrm{h}$ from the measurement of polarizability is only slightly higher \cite{huardBoundStatesOptical2000}. 




With increased carrier density, the temperatures concluded from the ODMR measurements -- also from the measurements with different magnetic field angles -- are higher than the temperatures derived from the giant Zeeman splitting with the same conditions. As mentioned before, presumably, a group of ions in the sample is somehow separated and needs a longer time to relax (reorient itself in the magnetic field) after being excited by a microwave pulse, but also to transfer energy in the form of temperature. This population would thus be more active in magnetic resonance signals. The rest would be less active in resonance studies, as they return to equilibrium much faster and their lines are significantly broadened. Nevertheless, this difference is observed only in the presence of carriers. For increased carrier density, the ions contributing to the ODMR signal thermalize in the ambient of hot carriers, which are excited by MW radiation non-resonantly. 

The influence of interaction with carriers on the ODMR resonance field depending on the angle of the magnetic field has already been analyzed in the works \cite{dietlZenerModelDescription2000,gurinODMREvidenceElectron2015}. In those works, the resonance field increases with increasing magnetic angle, which was explained by the interaction of a single, localized carrier with more than one \Mn\, ion -- which may be related to the fact that the concentration of manganese ions in these samples was higher $\sim$1 -2\% and the system was probably already in the magnetic polaron regime (Chapter 7 in Ref.\onlinecite{gajIntroductionPhysicsDiluted2010},\cite{godejohann2022trion, kavokin1999exciton}). In the case of the results presented here, the concentration of manganese ions is lower $\sim$0.3\%, and the model of delocalized carrier gas interacting with a multitude of \Mn\ ions is apparently more appropriate here \cite{hauryObservationFerromagneticTransition1997,dietlZenerModelDescription2000,kossackiLightControlledProbed2002, boukariLightElectricField2002}.

\section{Summary}
In this work, we showed that the ODMR spectra can provide information about the temperature of the magnetic ions and interacting carrier density. 

We simultaneously observe the ODMR signal as a change of the neutral and charged excitonic lines. Those two have different characteristics, proving the inhomogeneity of the hole gas distribution in the QW plane. 

Moreover, we found that in the presence of holes, the temperature of the magnetic ions contributing to the ODMR signal is slightly different than the temperature of the ions contributing to building net magnetization. That suggests the existence of two separated ensembles of magnetic ions, which are thermalized differently in the presence of carriers excited by microwave radiation.


\section{Acknowledgements}
This work was supported by the Polish National
Science Centre under decisions DEC-2016/23/B/ST3/03437,~DEC-2021/41/B/ST3/04183,~DEC-2020/38/E/ST3/00364,~DEC-2020/39/B/ST3/03251.


\begin{thebibliography}{31}
\expandafter\ifx\csname natexlab\endcsname\relax\def\natexlab#1{#1}\fi
\expandafter\ifx\csname bibnamefont\endcsname\relax
  \def\bibnamefont#1{#1}\fi
\expandafter\ifx\csname bibfnamefont\endcsname\relax
  \def\bibfnamefont#1{#1}\fi
\expandafter\ifx\csname citenamefont\endcsname\relax
  \def\citenamefont#1{#1}\fi
\expandafter\ifx\csname url\endcsname\relax
  \def\url#1{\texttt{#1}}\fi
\expandafter\ifx\csname urlprefix\endcsname\relax\def\urlprefix{URL }\fi
\providecommand{\bibinfo}[2]{#2}
\providecommand{\eprint}[2][]{\url{#2}}

\bibitem[{\citenamefont{Dietl and Ohno}(2014)}]{dietl2014dilute}
\bibinfo{author}{\bibfnamefont{T.}~\bibnamefont{Dietl}} \bibnamefont{and} \bibinfo{author}{\bibfnamefont{H.}~\bibnamefont{Ohno}}, \emph{\bibinfo{title}{Dilute ferromagnetic semiconductors: Physics and spintronic structures}}, \bibinfo{journal}{Reviews of Modern Physics} \textbf{\bibinfo{volume}{86}}, \bibinfo{pages}{187} (\bibinfo{year}{2014}).

\bibitem[{\citenamefont{Gaj and Kossut}(2010)}]{gajIntroductionPhysicsDiluted2010}
\bibinfo{editor}{\bibfnamefont{J.~A.} \bibnamefont{Gaj}} \bibnamefont{and} \bibinfo{editor}{\bibfnamefont{J.}~\bibnamefont{Kossut}}, eds., \emph{\bibinfo{title}{Introduction to the {{Physics}} of {{Diluted Magnetic Semiconductors}}}}, Springer {{Series}} in {{Materials Science}} (\bibinfo{publisher}{{Springer-Verlag}}, \bibinfo{address}{{Berlin Heidelberg}}, \bibinfo{year}{2010}), ISBN \bibinfo{isbn}{978-3-642-15855-1}.

\bibitem[{\citenamefont{Kossacki et~al.}(1999)\citenamefont{Kossacki, Cibert, Ferrand, {d'Aubign{\'e}}, Arnoult, Wasiela, Tatarenko, and Gaj}}]{kossackiNeutralPositivelyCharged1999}
\bibinfo{author}{\bibfnamefont{P.}~\bibnamefont{Kossacki}}, \bibinfo{author}{\bibfnamefont{J.}~\bibnamefont{Cibert}}, \bibinfo{author}{\bibfnamefont{D.}~\bibnamefont{Ferrand}}, \bibinfo{author}{\bibfnamefont{Y.~M.} \bibnamefont{{d'Aubign{\'e}}}}, \bibinfo{author}{\bibfnamefont{A.}~\bibnamefont{Arnoult}}, \bibinfo{author}{\bibfnamefont{A.}~\bibnamefont{Wasiela}}, \bibinfo{author}{\bibfnamefont{S.}~\bibnamefont{Tatarenko}}, \bibnamefont{and} \bibinfo{author}{\bibfnamefont{{\relax JA}.}~\bibnamefont{Gaj}}, \emph{\bibinfo{title}{Neutral and Positively Charged Excitons: A Magneto-Optical Study of a p-Doped {Cd}$_{(1-x)}${{Mg}}$_{x}$Te Quantum Well}}, \bibinfo{journal}{Physical Review B} \textbf{\bibinfo{volume}{60}}, \bibinfo{pages}{16018} (\bibinfo{year}{1999}).

\bibitem[{\citenamefont{Ma{\'s}lana et~al.}(2003)\citenamefont{Ma{\'s}lana, Kossacki, Bertolini, Boukari, Ferrand, Tatarenko, Cibert, and Gaj}}]{maslanaPtypeDopingII2003}
\bibinfo{author}{\bibfnamefont{W.}~\bibnamefont{Ma{\'s}lana}}, \bibinfo{author}{\bibfnamefont{P.}~\bibnamefont{Kossacki}}, \bibinfo{author}{\bibfnamefont{M.}~\bibnamefont{Bertolini}}, \bibinfo{author}{\bibfnamefont{H.}~\bibnamefont{Boukari}}, \bibinfo{author}{\bibfnamefont{D.}~\bibnamefont{Ferrand}}, \bibinfo{author}{\bibfnamefont{S.}~\bibnamefont{Tatarenko}}, \bibinfo{author}{\bibfnamefont{J.}~\bibnamefont{Cibert}}, \bibnamefont{and} \bibinfo{author}{\bibfnamefont{J.~A.} \bibnamefont{Gaj}}, \emph{\bibinfo{title}{P-Type Doping of {{II}}\textendash{{VI}} Heterostructures from Surface States: Application to Ferromagnetic {Cd}$_{(1-x)}${{Mg}}$_{x}$Te Quantum Wells}}, \bibinfo{journal}{Applied Physics Letters} \textbf{\bibinfo{volume}{82}}, \bibinfo{pages}{1875} (\bibinfo{year}{2003}).

\bibitem[{\citenamefont{Kossacki}(2003)}]{kossackiOpticalStudiesCharged2003}
\bibinfo{author}{\bibfnamefont{P.}~\bibnamefont{Kossacki}}, \emph{\bibinfo{title}{Optical {{Studies}} of Charged Excitons in {{II-VI}} Semiconductor Quantum Wells}}, \bibinfo{journal}{Journal of Physics: Condensed Matter} \textbf{\bibinfo{volume}{15}}, \bibinfo{pages}{R471} (\bibinfo{year}{2003}).

\bibitem[{\citenamefont{Kossacki et~al.}(2004)\citenamefont{Kossacki, Boukari, Bertolini, Ferrand, Cibert, Tatarenko, Gaj, Deveaud, Ciulin, and Potemski}}]{kossackiPhotoluminescencePDopedQuantum2004}
\bibinfo{author}{\bibfnamefont{P.}~\bibnamefont{Kossacki}}, \bibinfo{author}{\bibfnamefont{H.}~\bibnamefont{Boukari}}, \bibinfo{author}{\bibfnamefont{M.}~\bibnamefont{Bertolini}}, \bibinfo{author}{\bibfnamefont{D.}~\bibnamefont{Ferrand}}, \bibinfo{author}{\bibfnamefont{J.}~\bibnamefont{Cibert}}, \bibinfo{author}{\bibfnamefont{S.}~\bibnamefont{Tatarenko}}, \bibinfo{author}{\bibfnamefont{J.~A.} \bibnamefont{Gaj}}, \bibinfo{author}{\bibfnamefont{B.}~\bibnamefont{Deveaud}}, \bibinfo{author}{\bibfnamefont{V.}~\bibnamefont{Ciulin}}, \bibnamefont{and} \bibinfo{author}{\bibfnamefont{M.}~\bibnamefont{Potemski}}, \emph{\bibinfo{title}{Photoluminescence of P-Doped Quantum Wells with Strong Spin Splitting}}, \bibinfo{journal}{Physical Review B} \textbf{\bibinfo{volume}{70}}, \bibinfo{pages}{195337} (\bibinfo{year}{2004}).

\bibitem[{\citenamefont{{Aku-Leh} et~al.}(2007)\citenamefont{{Aku-Leh}, Perez, Jusserand, Richards, Pacuski, Kossacki, Menant, and Karczewski}}]{aku-lehMeasuringSpinPolarization2007}
\bibinfo{author}{\bibfnamefont{C.}~\bibnamefont{{Aku-Leh}}}, \bibinfo{author}{\bibfnamefont{F.}~\bibnamefont{Perez}}, \bibinfo{author}{\bibfnamefont{B.}~\bibnamefont{Jusserand}}, \bibinfo{author}{\bibfnamefont{D.}~\bibnamefont{Richards}}, \bibinfo{author}{\bibfnamefont{W.}~\bibnamefont{Pacuski}}, \bibinfo{author}{\bibfnamefont{P.}~\bibnamefont{Kossacki}}, \bibinfo{author}{\bibfnamefont{M.}~\bibnamefont{Menant}}, \bibnamefont{and} \bibinfo{author}{\bibfnamefont{G.}~\bibnamefont{Karczewski}}, \emph{\bibinfo{title}{Measuring the Spin Polarization and {{Zeeman}} Energy of a Spin-Polarized Electron Gas: Comparison between {{Raman}} Scattering and Photoluminescence}}, \bibinfo{journal}{Physical Review B} \textbf{\bibinfo{volume}{76}}, \bibinfo{pages}{155416} (\bibinfo{year}{2007}).

\bibitem[{\citenamefont{Baranov et~al.}(2008)\citenamefont{Baranov, Romanov, Tolmachev, Babunts, Namozov, Kusrayev, Sedova, Sorokin, and Ivanov}}]{baranov2008evidence}
\bibinfo{author}{\bibfnamefont{P.~G.} \bibnamefont{Baranov}}, \bibinfo{author}{\bibfnamefont{N.~G.} \bibnamefont{Romanov}}, \bibinfo{author}{\bibfnamefont{D.~O.} \bibnamefont{Tolmachev}}, \bibinfo{author}{\bibfnamefont{R.~A.} \bibnamefont{Babunts}}, \bibinfo{author}{\bibfnamefont{B.}~\bibnamefont{Namozov}}, \bibinfo{author}{\bibfnamefont{Y.~G.} \bibnamefont{Kusrayev}}, \bibinfo{author}{\bibfnamefont{I.~V.} \bibnamefont{Sedova}}, \bibinfo{author}{\bibfnamefont{S.~V.} \bibnamefont{Sorokin}}, \bibnamefont{and} \bibinfo{author}{\bibfnamefont{S.~V.} \bibnamefont{Ivanov}}, \emph{\bibinfo{title}{Evidence for Mn$^{2+}$ fine structure in CdMnSe/ZnSe quantum dots caused by their low dimensionality}}, \bibinfo{journal}{JETP letters} \textbf{\bibinfo{volume}{88}}, \bibinfo{pages}{631} (\bibinfo{year}{2008}).

\bibitem[{\citenamefont{Gurin et~al.}(2015)\citenamefont{Gurin, Tolmachev, Romanov, Namozov, Baranov, Kusrayev, and Karczewski}}]{gurinODMREvidenceElectron2015}
\bibinfo{author}{\bibfnamefont{A.~S.} \bibnamefont{Gurin}}, \bibinfo{author}{\bibfnamefont{D.~O.} \bibnamefont{Tolmachev}}, \bibinfo{author}{\bibfnamefont{N.~G.} \bibnamefont{Romanov}}, \bibinfo{author}{\bibfnamefont{B.~R.} \bibnamefont{Namozov}}, \bibinfo{author}{\bibfnamefont{P.~G.} \bibnamefont{Baranov}}, \bibinfo{author}{\bibfnamefont{{\relax Yu}.~G.} \bibnamefont{Kusrayev}}, \bibnamefont{and} \bibinfo{author}{\bibfnamefont{G.}~\bibnamefont{Karczewski}}, \emph{\bibinfo{title}{{{ODMR}} Evidence of the Electron Cascade in Multiple Asymmetrical ({{CdMn}}){{Te}} Quantum Wells}}, \bibinfo{journal}{JETP Letters} \textbf{\bibinfo{volume}{102}}, \bibinfo{pages}{230} (\bibinfo{year}{2015}).

\bibitem[{\citenamefont{{\L}opion et~al.}(2022)\citenamefont{{\L}opion, Bogucki, Kra{\'s}nicki, Po{\l}czy{\'n}ska, Pacuski, Kazimierczuk, Golnik, and Kossacki}}]{lopionMagneticIonRelaxation2022}
\bibinfo{author}{\bibfnamefont{A.}~\bibnamefont{{\L}opion}}, \bibinfo{author}{\bibfnamefont{A.}~\bibnamefont{Bogucki}}, \bibinfo{author}{\bibfnamefont{W.}~\bibnamefont{Kra{\'s}nicki}}, \bibinfo{author}{\bibfnamefont{K.~E.} \bibnamefont{Po{\l}czy{\'n}ska}}, \bibinfo{author}{\bibfnamefont{W.}~\bibnamefont{Pacuski}}, \bibinfo{author}{\bibfnamefont{T.}~\bibnamefont{Kazimierczuk}}, \bibinfo{author}{\bibfnamefont{A.}~\bibnamefont{Golnik}}, \bibnamefont{and} \bibinfo{author}{\bibfnamefont{P.}~\bibnamefont{Kossacki}}, \emph{\bibinfo{title}{Magnetic Ion Relaxation Time Distribution within a Quantum Well}}, \bibinfo{journal}{Physical Review B} \textbf{\bibinfo{volume}{106}}, \bibinfo{pages}{165309} (\bibinfo{year}{2022}).

\bibitem[{\citenamefont{Waag et~al.}(1993)\citenamefont{Waag, Heinke, Scholl, Becker, and Landwehr}}]{waagGrowthMgTeCd1993}
\bibinfo{author}{\bibfnamefont{A.}~\bibnamefont{Waag}}, \bibinfo{author}{\bibfnamefont{H.}~\bibnamefont{Heinke}}, \bibinfo{author}{\bibfnamefont{S.}~\bibnamefont{Scholl}}, \bibinfo{author}{\bibfnamefont{C.~R.} \bibnamefont{Becker}}, \bibnamefont{and} \bibinfo{author}{\bibfnamefont{G.}~\bibnamefont{Landwehr}}, \emph{\bibinfo{title}{Growth of {{MgTe}} and {{Cd}}$_{(1-x)}${{Mg}}$_{x}$Te Thin Films by {{Molecular Beam Epitaxy}}}}, \bibinfo{journal}{Journal of Crystal Growth} \textbf{\bibinfo{volume}{131}}, \bibinfo{pages}{607} (\bibinfo{year}{1993}).

\bibitem[{\citenamefont{Gaj et~al.}(1994)\citenamefont{Gaj, Grieshaber, {Bodin-Deshayes}, Cibert, Feuillet, {Merle d'Aubign{\'e}}, and Wasiela}}]{gajMagnetoopticalStudyInterface1994}
\bibinfo{author}{\bibfnamefont{J.~A.} \bibnamefont{Gaj}}, \bibinfo{author}{\bibfnamefont{W.}~\bibnamefont{Grieshaber}}, \bibinfo{author}{\bibfnamefont{C.}~\bibnamefont{{Bodin-Deshayes}}}, \bibinfo{author}{\bibfnamefont{J.}~\bibnamefont{Cibert}}, \bibinfo{author}{\bibfnamefont{G.}~\bibnamefont{Feuillet}}, \bibinfo{author}{\bibfnamefont{Y.}~\bibnamefont{{Merle d'Aubign{\'e}}}}, \bibnamefont{and} \bibinfo{author}{\bibfnamefont{A.}~\bibnamefont{Wasiela}}, \emph{\bibinfo{title}{Magneto-Optical Study of Interface Mixing in the {{CdTe-}}({{Cd}},{{Mn}}){{Te}} System}}, \bibinfo{journal}{Physical Review B} \textbf{\bibinfo{volume}{50}}, \bibinfo{pages}{5512} (\bibinfo{year}{1994}).

\bibitem[{\citenamefont{{\L}opion et~al.}(2020)\citenamefont{{\L}opion, Bogucki, Po{\l}czy{\'n}ska, Pacuski, Golnik, Kazimierczuk, and Kossacki}}]{lopionChargedExcitonDissociation2020}
\bibinfo{author}{\bibfnamefont{A.}~\bibnamefont{{\L}opion}}, \bibinfo{author}{\bibfnamefont{A.}~\bibnamefont{Bogucki}}, \bibinfo{author}{\bibfnamefont{K.~E.} \bibnamefont{Po{\l}czy{\'n}ska}}, \bibinfo{author}{\bibfnamefont{W.}~\bibnamefont{Pacuski}}, \bibinfo{author}{\bibfnamefont{A.}~\bibnamefont{Golnik}}, \bibinfo{author}{\bibfnamefont{T.}~\bibnamefont{Kazimierczuk}}, \bibnamefont{and} \bibinfo{author}{\bibfnamefont{P.}~\bibnamefont{Kossacki}}, \emph{\bibinfo{title}{Charged Exciton Dissociation Energy in ({{Cd}},{{Mn}}){{Te}} Quantum Wells with Variable Disorder and Carrier Density}}, \bibinfo{journal}{Journal of Electronic Materials}  (\bibinfo{year}{2020}).

\bibitem[{\citenamefont{Huard et~al.}(2000)\citenamefont{Huard, Cox, Saminadayar, Arnoult, and Tatarenko}}]{huardBoundStatesOptical2000}
\bibinfo{author}{\bibfnamefont{V.}~\bibnamefont{Huard}}, \bibinfo{author}{\bibfnamefont{{\relax RT}.}~\bibnamefont{Cox}}, \bibinfo{author}{\bibfnamefont{K.}~\bibnamefont{Saminadayar}}, \bibinfo{author}{\bibfnamefont{A.}~\bibnamefont{Arnoult}}, \bibnamefont{and} \bibinfo{author}{\bibfnamefont{S.}~\bibnamefont{Tatarenko}}, \emph{\bibinfo{title}{Bound States in Optical Absorption of Semiconductor Quantum Wells Containing a Two-Dimensional Electron Gas}}, \bibinfo{journal}{Physical Review Letters} \textbf{\bibinfo{volume}{84}}, \bibinfo{pages}{187} (\bibinfo{year}{2000}).

\bibitem[{\citenamefont{Bogucki et~al.}(2022)\citenamefont{Bogucki, Goryca, {\L}opion, Pacuski, Po{\l}czy{\'n}ska, Domaga{\l}a, Tokarczyk, F{\k{a}}s, Golnik, and Kossacki}}]{boguckiAngleresolvedOpticallyDetected2022}
\bibinfo{author}{\bibfnamefont{A.}~\bibnamefont{Bogucki}}, \bibinfo{author}{\bibfnamefont{M.}~\bibnamefont{Goryca}}, \bibinfo{author}{\bibfnamefont{A.}~\bibnamefont{{\L}opion}}, \bibinfo{author}{\bibfnamefont{W.}~\bibnamefont{Pacuski}}, \bibinfo{author}{\bibfnamefont{K.}~\bibnamefont{Po{\l}czy{\'n}ska}}, \bibinfo{author}{\bibfnamefont{J.}~\bibnamefont{Domaga{\l}a}}, \bibinfo{author}{\bibfnamefont{M.}~\bibnamefont{Tokarczyk}}, \bibinfo{author}{\bibfnamefont{T.}~\bibnamefont{F{\k{a}}s}}, \bibinfo{author}{\bibfnamefont{A.}~\bibnamefont{Golnik}}, \bibnamefont{and} \bibinfo{author}{\bibfnamefont{P.}~\bibnamefont{Kossacki}}, \emph{\bibinfo{title}{Angle-Resolved Optically Detected Magnetic Resonance as a Tool for Strain Determination in Nanostructures}}, \bibinfo{journal}{Physical Review B} \textbf{\bibinfo{volume}{105}}, \bibinfo{pages}{075412} (\bibinfo{year}{2022}).

\bibitem[{\citenamefont{Qazzaz et~al.}(1995)\citenamefont{Qazzaz, Yang, Xin, Montes, Luo, and Furdyna}}]{qazzazElectronParamagneticResonance1995}
\bibinfo{author}{\bibfnamefont{M.}~\bibnamefont{Qazzaz}}, \bibinfo{author}{\bibfnamefont{G.}~\bibnamefont{Yang}}, \bibinfo{author}{\bibfnamefont{S.~H.} \bibnamefont{Xin}}, \bibinfo{author}{\bibfnamefont{L.}~\bibnamefont{Montes}}, \bibinfo{author}{\bibfnamefont{H.}~\bibnamefont{Luo}}, \bibnamefont{and} \bibinfo{author}{\bibfnamefont{J.~K.} \bibnamefont{Furdyna}}, \emph{\bibinfo{title}{Electron Paramagnetic Resonance of Mn$^{2+}$ in Strained-Layer Semiconductor Superlattices}}, \bibinfo{journal}{Solid State Communications} \textbf{\bibinfo{volume}{96}}, \bibinfo{pages}{405} (\bibinfo{year}{1995}).

\bibitem[{\citenamefont{Abragam}(1961)}]{abragamPrinciplesNuclearMagnetism1961}
\bibinfo{author}{\bibfnamefont{A.}~\bibnamefont{Abragam}}, \emph{\bibinfo{title}{The {{Principles}} of {{Nuclear Magnetism}}}} (\bibinfo{publisher}{{Clarendon Press}}, \bibinfo{year}{1961}), ISBN \bibinfo{isbn}{978-0-19-852014-6}.

\bibitem[{\citenamefont{Greenough and Palmer}(1973)}]{greenoughElasticConstantsThermal1973}
\bibinfo{author}{\bibfnamefont{R.~D.} \bibnamefont{Greenough}} \bibnamefont{and} \bibinfo{author}{\bibfnamefont{S.~B.} \bibnamefont{Palmer}}, \emph{\bibinfo{title}{The Elastic Constants and Thermal Expansion of Single-Crystal {{CdTe}}}}, \bibinfo{journal}{Journal of Physics D: Applied Physics} \textbf{\bibinfo{volume}{6}}, \bibinfo{pages}{587} (\bibinfo{year}{1973}).

\bibitem[{\citenamefont{Golnik et~al.}(2004)\citenamefont{Golnik, Kossacki, Kowalik, Ma{\'s}lana, Gaj, Kutrowski, and Wojtowicz}}]{golnikMicrophotoluminescenceStudyLocal2004}
\bibinfo{author}{\bibfnamefont{A.}~\bibnamefont{Golnik}}, \bibinfo{author}{\bibfnamefont{P.}~\bibnamefont{Kossacki}}, \bibinfo{author}{\bibfnamefont{K.}~\bibnamefont{Kowalik}}, \bibinfo{author}{\bibfnamefont{W.}~\bibnamefont{Ma{\'s}lana}}, \bibinfo{author}{\bibfnamefont{{\relax JA}.}~\bibnamefont{Gaj}}, \bibinfo{author}{\bibfnamefont{M.}~\bibnamefont{Kutrowski}}, \bibnamefont{and} \bibinfo{author}{\bibfnamefont{T.}~\bibnamefont{Wojtowicz}}, \emph{\bibinfo{title}{Microphotoluminescence Study of Local Temperature Fluctuations in N-Type ({{Cd}}, {{Mn}}) {{Te}} Quantum Well}}, \bibinfo{journal}{Solid State Communications} \textbf{\bibinfo{volume}{131}}, \bibinfo{pages}{283} (\bibinfo{year}{2004}).

\bibitem[{\citenamefont{Ma{\'s}lana et~al.}(2006)\citenamefont{Ma{\'s}lana, Kossacki, P{\l}ochocka, Golnik, Gaj, Ferrand, Bertolini, Tatarenko, and Cibert}}]{MaslanaMicrophotoluminescenceStudyPtype2006}
\bibinfo{author}{\bibfnamefont{W.}~\bibnamefont{Ma{\'s}lana}}, \bibinfo{author}{\bibfnamefont{P.}~\bibnamefont{Kossacki}}, \bibinfo{author}{\bibfnamefont{P.}~\bibnamefont{P{\l}ochocka}}, \bibinfo{author}{\bibfnamefont{A.}~\bibnamefont{Golnik}}, \bibinfo{author}{\bibfnamefont{{\relax JA}.}~\bibnamefont{Gaj}}, \bibinfo{author}{\bibfnamefont{D.}~\bibnamefont{Ferrand}}, \bibinfo{author}{\bibfnamefont{M.}~\bibnamefont{Bertolini}}, \bibinfo{author}{\bibfnamefont{S.}~\bibnamefont{Tatarenko}}, \bibnamefont{and} \bibinfo{author}{\bibfnamefont{J.}~\bibnamefont{Cibert}}, \emph{\bibinfo{title}{Microphotoluminescence Study of P-Type ({{Cd}}, {{Mn}}){{Te}} Quantum Wells}}, \bibinfo{journal}{Applied Physics Letters} \textbf{\bibinfo{volume}{89}}, \bibinfo{pages}{052104} (\bibinfo{year}{2006}).

\bibitem[{\citenamefont{Moody et~al.}(2014)\citenamefont{Moody, Akimov, Li, Singh, Yakovlev, Karczewski, Wiater, Wojtowicz, Bayer, and Cundiff}}]{moody2014coherent}
\bibinfo{author}{\bibfnamefont{G.}~\bibnamefont{Moody}}, \bibinfo{author}{\bibfnamefont{I.}~\bibnamefont{Akimov}}, \bibinfo{author}{\bibfnamefont{H.}~\bibnamefont{Li}}, \bibinfo{author}{\bibfnamefont{R.}~\bibnamefont{Singh}}, \bibinfo{author}{\bibfnamefont{D.}~\bibnamefont{Yakovlev}}, \bibinfo{author}{\bibfnamefont{G.}~\bibnamefont{Karczewski}}, \bibinfo{author}{\bibfnamefont{M.}~\bibnamefont{Wiater}}, \bibinfo{author}{\bibfnamefont{T.}~\bibnamefont{Wojtowicz}}, \bibinfo{author}{\bibfnamefont{M.}~\bibnamefont{Bayer}}, \bibnamefont{and} \bibinfo{author}{\bibfnamefont{S.}~\bibnamefont{Cundiff}}, \emph{\bibinfo{title}{Coherent coupling of excitons and trions in a photoexcited CdTe/CdMgTe quantum well}}, \bibinfo{journal}{Physical review letters} \textbf{\bibinfo{volume}{112}}, \bibinfo{pages}{097401} (\bibinfo{year}{2014}).

\bibitem[{\citenamefont{Brunhes et~al.}(1999)\citenamefont{Brunhes, Andr{\'e}, Arnoult, Cibert, and Wasiela}}]{brunhes1999oscillator}
\bibinfo{author}{\bibfnamefont{T.}~\bibnamefont{Brunhes}}, \bibinfo{author}{\bibfnamefont{R.}~\bibnamefont{Andr{\'e}}}, \bibinfo{author}{\bibfnamefont{A.}~\bibnamefont{Arnoult}}, \bibinfo{author}{\bibfnamefont{J.}~\bibnamefont{Cibert}}, \bibnamefont{and} \bibinfo{author}{\bibfnamefont{A.}~\bibnamefont{Wasiela}}, \emph{\bibinfo{title}{Oscillator strength transfer from X to X$^+$ in a CdTe quantum-well microcavity}}, \bibinfo{journal}{Physical Review B} \textbf{\bibinfo{volume}{60}}, \bibinfo{pages}{11568} (\bibinfo{year}{1999}).

\bibitem[{\citenamefont{Scherbakov et~al.}(2001)\citenamefont{Scherbakov, Yakovlev, Akimov, Merkulov, K{\"o}nig, Ossau, Molenkamp, Wojtowicz, Karczewski, Cywinski et~al.}}]{scherbakovAccelerationSpinlatticeRelaxation2001}
\bibinfo{author}{\bibfnamefont{{\relax AV}.}~\bibnamefont{Scherbakov}}, \bibinfo{author}{\bibfnamefont{{\relax DR}.}~\bibnamefont{Yakovlev}}, \bibinfo{author}{\bibfnamefont{{\relax AV}.}~\bibnamefont{Akimov}}, \bibinfo{author}{\bibfnamefont{{\relax IA}.}~\bibnamefont{Merkulov}}, \bibinfo{author}{\bibfnamefont{B.}~\bibnamefont{K{\"o}nig}}, \bibinfo{author}{\bibfnamefont{W.}~\bibnamefont{Ossau}}, \bibinfo{author}{\bibfnamefont{{\relax LW}.}~\bibnamefont{Molenkamp}}, \bibinfo{author}{\bibfnamefont{T.}~\bibnamefont{Wojtowicz}}, \bibinfo{author}{\bibfnamefont{G.}~\bibnamefont{Karczewski}}, \bibinfo{author}{\bibfnamefont{G.}~\bibnamefont{Cywinski}}, \bibnamefont{and} \bibinfo{author}{\bibnamefont{{Kossut, J}}}, \emph{\bibinfo{title}{Acceleration of the Spin-Lattice Relaxation in Diluted Magnetic Quantum Wells in the Presence of a Two-Dimensional Electron Gas}}, \bibinfo{journal}{Physical Review B} \textbf{\bibinfo{volume}{64}}, \bibinfo{pages}{155205} (\bibinfo{year}{2001}).

\bibitem[{\citenamefont{K{\"o}nig et~al.}(2000)\citenamefont{K{\"o}nig, Merkulov, Yakovlev, Ossau, Ryabchenko, Kutrowski, Wojtowicz, Karczewski, and Kossut}}]{konigEnergyTransferPhotocarriers2000}
\bibinfo{author}{\bibfnamefont{B.}~\bibnamefont{K{\"o}nig}}, \bibinfo{author}{\bibfnamefont{I.~A.} \bibnamefont{Merkulov}}, \bibinfo{author}{\bibfnamefont{D.~R.} \bibnamefont{Yakovlev}}, \bibinfo{author}{\bibfnamefont{W.}~\bibnamefont{Ossau}}, \bibinfo{author}{\bibfnamefont{S.~M.} \bibnamefont{Ryabchenko}}, \bibinfo{author}{\bibfnamefont{M.}~\bibnamefont{Kutrowski}}, \bibinfo{author}{\bibfnamefont{T.}~\bibnamefont{Wojtowicz}}, \bibinfo{author}{\bibfnamefont{G.}~\bibnamefont{Karczewski}}, \bibnamefont{and} \bibinfo{author}{\bibfnamefont{J.}~\bibnamefont{Kossut}}, \emph{\bibinfo{title}{Energy Transfer from Photocarriers into the Magnetic Ion System Mediated by a Two-Dimensional Electron Gas in ({{Cd}},{{Mn}}){{Te}}/({{Cd}},{{Mg}}){{Te}} Quantum Wells}}, \bibinfo{journal}{Physical Review B} \textbf{\bibinfo{volume}{61}}, \bibinfo{pages}{16870} (\bibinfo{year}{2000}).

\bibitem[{\citenamefont{Scherbakov et~al.}(1999)\citenamefont{Scherbakov, Akimov, Yakovlev, Ossau, Waag, Landwehr, Wojtowicz, Karczewski, and Kossut}}]{scherbakov1999heating}
\bibinfo{author}{\bibfnamefont{A.}~\bibnamefont{Scherbakov}}, \bibinfo{author}{\bibfnamefont{A.}~\bibnamefont{Akimov}}, \bibinfo{author}{\bibfnamefont{D.}~\bibnamefont{Yakovlev}}, \bibinfo{author}{\bibfnamefont{W.}~\bibnamefont{Ossau}}, \bibinfo{author}{\bibfnamefont{A.}~\bibnamefont{Waag}}, \bibinfo{author}{\bibfnamefont{G.}~\bibnamefont{Landwehr}}, \bibinfo{author}{\bibfnamefont{T.}~\bibnamefont{Wojtowicz}}, \bibinfo{author}{\bibfnamefont{G.}~\bibnamefont{Karczewski}}, \bibnamefont{and} \bibinfo{author}{\bibfnamefont{J.}~\bibnamefont{Kossut}}, \emph{\bibinfo{title}{Heating of the spin system by nonequilibrium phonons in semimagnetic (Cd, Mn, Mg) Te quantum wells}}, \bibinfo{journal}{Physical Review B} \textbf{\bibinfo{volume}{60}}, \bibinfo{pages}{5609} (\bibinfo{year}{1999}).

\bibitem[{\citenamefont{Dietl et~al.}(2000)\citenamefont{Dietl, Ohno, Matsukura, Cibert, and Ferrand}}]{dietlZenerModelDescription2000}
\bibinfo{author}{\bibfnamefont{T.}~\bibnamefont{Dietl}}, \bibinfo{author}{\bibfnamefont{o.~H.} \bibnamefont{Ohno}}, \bibinfo{author}{\bibfnamefont{a.~F.} \bibnamefont{Matsukura}}, \bibinfo{author}{\bibfnamefont{J.}~\bibnamefont{Cibert}}, \bibnamefont{and} \bibinfo{author}{\bibfnamefont{e.~D.} \bibnamefont{Ferrand}}, \emph{\bibinfo{title}{Zener Model Description of Ferromagnetism in Zinc-Blende Magnetic Semiconductors}}, \bibinfo{journal}{Science} \textbf{\bibinfo{volume}{287}}, \bibinfo{pages}{1019} (\bibinfo{year}{2000}).

\bibitem[{\citenamefont{Godejohann et~al.}(2022)\citenamefont{Godejohann, Akhmadullin, Kavokin, Yakovlev, Akimov, Namozov, Kusrayev, Karczewski, Wojtowicz, and Bayer}}]{godejohann2022trion}
\bibinfo{author}{\bibfnamefont{F.}~\bibnamefont{Godejohann}}, \bibinfo{author}{\bibfnamefont{R.}~\bibnamefont{Akhmadullin}}, \bibinfo{author}{\bibfnamefont{K.}~\bibnamefont{Kavokin}}, \bibinfo{author}{\bibfnamefont{D.}~\bibnamefont{Yakovlev}}, \bibinfo{author}{\bibfnamefont{I.}~\bibnamefont{Akimov}}, \bibinfo{author}{\bibfnamefont{B.}~\bibnamefont{Namozov}}, \bibinfo{author}{\bibfnamefont{Y.~G.} \bibnamefont{Kusrayev}}, \bibinfo{author}{\bibfnamefont{G.}~\bibnamefont{Karczewski}}, \bibinfo{author}{\bibfnamefont{T.}~\bibnamefont{Wojtowicz}}, \bibnamefont{and} \bibinfo{author}{\bibfnamefont{M.}~\bibnamefont{Bayer}}, \emph{\bibinfo{title}{Trion magnetic polarons in (Cd, Mn) Te/(Cd, Mn, Mg) Te quantum wells}}, \bibinfo{journal}{Physical Review B} \textbf{\bibinfo{volume}{106}}, \bibinfo{pages}{195305} (\bibinfo{year}{2022}).

\bibitem[{\citenamefont{Kavokin et~al.}(1999)\citenamefont{Kavokin, Merkulov, Yakovlev, Ossau, and Landwehr}}]{kavokin1999exciton}
\bibinfo{author}{\bibfnamefont{K.}~\bibnamefont{Kavokin}}, \bibinfo{author}{\bibfnamefont{I.}~\bibnamefont{Merkulov}}, \bibinfo{author}{\bibfnamefont{D.}~\bibnamefont{Yakovlev}}, \bibinfo{author}{\bibfnamefont{W.}~\bibnamefont{Ossau}}, \bibnamefont{and} \bibinfo{author}{\bibfnamefont{G.}~\bibnamefont{Landwehr}}, \emph{\bibinfo{title}{Exciton localization in semimagnetic semiconductors probed by magnetic polarons}}, \bibinfo{journal}{Physical Review B} \textbf{\bibinfo{volume}{60}}, \bibinfo{pages}{16499} (\bibinfo{year}{1999}).

\bibitem[{\citenamefont{Haury et~al.}(1997)\citenamefont{Haury, Wasiela, Arnoult, Cibert, Tatarenko, Dietl, and {d'Aubign{\'e}}}}]{hauryObservationFerromagneticTransition1997}
\bibinfo{author}{\bibfnamefont{A.}~\bibnamefont{Haury}}, \bibinfo{author}{\bibfnamefont{A.}~\bibnamefont{Wasiela}}, \bibinfo{author}{\bibfnamefont{t.~A.} \bibnamefont{Arnoult}}, \bibinfo{author}{\bibfnamefont{J.}~\bibnamefont{Cibert}}, \bibinfo{author}{\bibfnamefont{S.}~\bibnamefont{Tatarenko}}, \bibinfo{author}{\bibfnamefont{T.}~\bibnamefont{Dietl}}, \bibnamefont{and} \bibinfo{author}{\bibfnamefont{Y.~M.} \bibnamefont{{d'Aubign{\'e}}}}, \emph{\bibinfo{title}{Observation of a Ferromagnetic Transition Induced by Two-Dimensional Hole Gas in Modulation-Doped {{CdMnTe}} Quantum Wells}}, \bibinfo{journal}{Physical Review Letters} \textbf{\bibinfo{volume}{79}}, \bibinfo{pages}{511} (\bibinfo{year}{1997}).

\bibitem[{\citenamefont{Kossacki et~al.}(2002)\citenamefont{Kossacki, Kudelski, Gaj, Cibert, Tatarenko, Ferrand, Wasiela, Deveaud, and Dietl}}]{kossackiLightControlledProbed2002}
\bibinfo{author}{\bibfnamefont{P.}~\bibnamefont{Kossacki}}, \bibinfo{author}{\bibfnamefont{A.}~\bibnamefont{Kudelski}}, \bibinfo{author}{\bibfnamefont{J.}~\bibnamefont{Gaj}}, \bibinfo{author}{\bibfnamefont{J.}~\bibnamefont{Cibert}}, \bibinfo{author}{\bibfnamefont{S.}~\bibnamefont{Tatarenko}}, \bibinfo{author}{\bibfnamefont{D.}~\bibnamefont{Ferrand}}, \bibinfo{author}{\bibfnamefont{A.}~\bibnamefont{Wasiela}}, \bibinfo{author}{\bibfnamefont{B.}~\bibnamefont{Deveaud}}, \bibnamefont{and} \bibinfo{author}{\bibfnamefont{T.}~\bibnamefont{Dietl}}, \emph{\bibinfo{title}{Light {{Controlled}} and {{Probed Ferromagnetism}} of ({{Cd}},{{Mn}}){{Te Quantum Wells}}}}, \bibinfo{journal}{Physica E: Low-dimensional Systems and Nanostructures} \textbf{\bibinfo{volume}{12}}, \bibinfo{pages}{344} (\bibinfo{year}{2002}).

\bibitem[{\citenamefont{Boukari et~al.}(2002)\citenamefont{Boukari, Kossacki, Bertolini, Ferrand, Cibert, Tatarenko, Wasiela, Gaj, and Dietl}}]{boukariLightElectricField2002}
\bibinfo{author}{\bibfnamefont{H.}~\bibnamefont{Boukari}}, \bibinfo{author}{\bibfnamefont{P.}~\bibnamefont{Kossacki}}, \bibinfo{author}{\bibfnamefont{M.}~\bibnamefont{Bertolini}}, \bibinfo{author}{\bibfnamefont{D.}~\bibnamefont{Ferrand}}, \bibinfo{author}{\bibfnamefont{J.}~\bibnamefont{Cibert}}, \bibinfo{author}{\bibfnamefont{S.}~\bibnamefont{Tatarenko}}, \bibinfo{author}{\bibfnamefont{A.}~\bibnamefont{Wasiela}}, \bibinfo{author}{\bibfnamefont{{\relax JA}.}~\bibnamefont{Gaj}}, \bibnamefont{and} \bibinfo{author}{\bibfnamefont{T.}~\bibnamefont{Dietl}}, \emph{\bibinfo{title}{Light and Electric Field Control of Ferromagnetism in Magnetic Quantum Structures}}, \bibinfo{journal}{Physical Review Letters} \textbf{\bibinfo{volume}{88}}, \bibinfo{pages}{207204} (\bibinfo{year}{2002}).

\end{thebibliography}
\end{document}